\documentclass[10pt,showpacs,bibnotes,notitlepage,twocolumn,balancelastpage,preprintnumbers,amsmath,amssymb,final,pra]{revtex4}
\usepackage{graphicx}
 \usepackage{color}

\begin{document}

\title{Two-species hard-core bosons on the triangular lattice: A quantum Monte Carlo study}
\author{Jian-Ping Lv$^{1,2}$}
\email{phys.lv@gmail.com}
\author{Qing-Hu Chen$^{2,3}$}
\email{qhchen@zju.edu.cn}
\author{Youjin Deng$^{4}$}
\email{yjdeng@ustc.edu.cn}

\address{
$^{1}$  Department of Physics, China University of Mining and
Technology, Xuzhou 221116, P. R. China \\
$^{2}$ Department of
Physics, Zhejiang University, Hangzhou 310027,
P. R. China \\
$^{3}$ Center for Statistical and Theoretical Condensed Matter
Physics, Zhejiang Normal University, Jinhua 321004, P. R. China \\
$^{4}$ Hefei National Laboratory for Physical Sciences at Microscale
and Department of Modern Physics, University of Science and
Technology of China, Hefei 230027, P. R. China}

\date{\today}

\begin{abstract}
Using worm-type quantum Monte Carlo simulations, we investigate
bosonic mixtures on the triangular lattice of two species of bosons, which interact
via nearest-neighbour intraspecies ($V$) and onsite interspecies ($U$) repulsions.
For the case of symmetric hopping amplitude ($t_A/V=t_B/V$) and $U/V=1$,
we determine a rich ground-state phase diagram that contains double solid, double superfluid (2SF),
supersolid (SS), solid-superfluid (Solid-SF) and counterflow supersolid (CSS) states.
The SS, Solid-SF and CSS states exhibit spontaneous symmetry breaking among the three
sublattices of the triangular lattice and between the two species,
which leads to nonzero crystalline density wave order in each species.
We furthermore show that the CSS and the SS states are present for $t_A/V \neq t_B/V$, and the latter even
survives up to $t_A/V \rightarrow \infty$ or $t_B/V \rightarrow \infty$ limit.
The effects induced by the variation of $U/V$ and by the imbalance of particle numbers
of the two species are also explored.
\end{abstract}
\maketitle
%%%
\section{Introduction}
Ultracold bosons held in optical lattices provide an ideal realization of the single-species Bose-Hubbard model~\cite{Greiner},
and attract extensive interest from both experimental and theoretical research
communities (for review, see e.g., Refs.~\cite{rev1} and \cite{rev2}).
Recently, more and more attention has been paid to the two-species bosons where
novel quantum phases can emerge due to interspecies and intraspecies
interactions~\cite{TC1,TC_m,Guglielmino2010,Hubener2009,TC4,Hettiarachchilage,ChenYang,BCS,He2012,Iskin,Trefzger2009,Trousselet2013,Pakrouski,Boninsegni,Kuklov,Kataoka2012,Kuno2013},
which can be tuned experimentally by Feshbach resonances~\cite{Catani,Thalhammer}.

The quantum phases and phase transitions in two-species bosons on bipartite lattices
with onsite interspecies interaction $(U)$ have been extensively studied.
 Kuklov and Svistunov demonstrated a novel quantum phase--the so-called counterflow superfluid (CSF)--and
constructed an effective Hamiltonian~\cite{Kuklov}.
The CSF state features nonzero CSF density but
vanishing pair superfluid (PSF) density between the two species.
Altman $et$ $al.$ investigated the hardcore case with each species at half-integer filling ($\rho_A=\rho_B=1/2$)
within a mean-field approach and established for $U>0$ a phase diagram on the $t_A/U$-$t_B/U$ plane~\cite{TC_m}.
The phase diagram contains CSF, checkerboard solid, and superfluid (SF) phases.
Using a worm-type quantum Monte Carlo method, S\"{o}yler $et$ $al.$~\cite{TC1}
found that in the strongly asymmetric region, the phase diagram for the square lattice differs
from the mean-field result~\cite{TC_m}.
More direct evidence for the emergence of the CSF phase can come from
measurement of pair-correlation functions~\cite{TC4}.
The robustness of these quantum ordered phases against thermodynamic fluctuations was then explored
by Capogrosso-Sansone $et$ $al.$~\cite{BCS}, and finite-temperature phase transitions were obtained
for the square and simple-cubic lattices.
For systems away from the half-integer fillings~\cite{Hettiarachchilage} or
with attractive onsite interspecies interaction $U<0$~\cite{ChenYang}, other quantum phases
can be found--e.g., the emergence of PSF phase in the latter.
Systems of two-species softcore bosons have also been studied~\cite{Guglielmino2010,Trefzger2009}.

In the past few years, significant experimental progress has been achieved and
ultracold atoms can be loaded into different optical lattices such as
the triangular~\cite{triOL}, kagom\'e~\cite{kagOL}, and dice~\cite{diceOL} lattices,
on which rich physics can occur due to distinct band structure or geometric frustration etc~\cite{tri2005,kagome}.
For the two-species softcore bosons on the triangular lattice with onsite intraspecies and
onsite interspecies interactions, a phase diagram has been established recently~\cite{He2012}.

In this work, we perform extensive Monte Carlo simulations on
bosonic mixtures on the triangular lattice, which are constituted by two species of hardcore bosons with both
onsite interspecies and nearest-neighbour intraspecies repulsions.
This model was recently studied by Trousselet $et$ $al.$ using a mean-filed approach combined with
exact diagnolizations~\cite{Trousselet2013}. The organization of this paper is as follows. Section~\ref{s2} introduces the model and presents analysis
for some limiting cases. Measured quantities are defined in Sec.~\ref{s4}.
Section~\ref{s5} describes numerical results, and a brief discussion is given in Sec.~\ref{s6}.

\section{Model}~\label{s2}

Let the two species of hardcore bosons be specified by $A$ and $B$ and the associated
creation (annihilation) operators be $a^{\dag}_i (a^{ }_i)$ and  $b^{\dag}_i(b^{}_i)$,
the Hamiltonian studied in this work can be written as
\begin{eqnarray}
H=&-&t_A\! \sum_{<ij>} a^+_i a_j -t_B\! \sum_{<ij>}  b^+_i b_j  \nonumber \\
  &+&  V \sum_{<ij>}(n^{A}_i  n^{A}_j+n^{B}_i  n^{B}_j) \nonumber \\
  &+ & U \sum_{i}  n^{A}_i  n^{B}_i-\mu \sum_{i}  (n^{A}_i  +n^{B}_i),
\label{hamiltonian}
\end{eqnarray}
where $t_A$ and $t_B$ are the hopping amplitudes, and $V$ and $U$ are
the nearest-neighbor intraspecies and onsite interspecies repulsions, respectively.
Symbols $n^{A}_i=a^{\dag}_i a^{ }_i$ and $n^{B}_i=b^{\dag}_i b^{ }_i$
are the particle-number operators.
Due to the hardcore constraint, one has
$\{a_i,a_i\}=\{a^+_i,a^+_i\}= \{b_i,b_i\}=\{b^+_i,b^+_i\}=0$ and
$\{a_i,a^+_i\}=\{b_i,b^+_i\}=1$.
The Hamiltonian is the same with that studied in Ref.~\cite{Trousselet2013} and
similar to that in Ref.~\cite{Trefzger2009}.

$\texttt{Particle-hole symmetry}$.  Employing the particle-hole transformations,
$U_A^{+} a U_A  = a^+$ and $U_B^{+} b U_B = b^+$, we have
\begin{equation}
U^{+}_{A} U^{+}_{B} H(\mu) U_B U_A=H(6V+U-\mu).
\label{ph}
\end{equation}
Therefore, Model~(\ref{hamiltonian}) exhibits an exact particle-hole symmetry at $\mu=3V+U/2$.
It follows that, to simulate the model with half-integer filling factor of each species, one can perform grand-canonical simulations with $\mu=3V+U/2$.
This treatment was elaborated in Ref.~\cite{Pakrouski} and also employed in a dynamical mean-field study
of bosonic mixtures~\cite{Hubener2009}.

$\texttt{Classical limit  ($t_A$/V=$t_B$/V=0)}$. In the zero-hopping limit $t_A/V=t_B/V=0$,
 Hamiltonian~(\ref{hamiltonian}) reduces to
\begin{eqnarray}
H = &V& \sum_{<ij>}(n^{A}_i  n^{A}_j+n^{B}_i  n^{B}_j) + U \sum_{i}  n^{A}_i  n^{B}_i \nonumber \\
   &-&\mu \sum_{i}  (n^{A}_i  +n^{B}_i).
\label{hamiltonian01}
\end{eqnarray}
At zero temperature $T=0$, thermal fluctuations are frozen, and
there are two possible solid states:
\begin{enumerate}
\item 2Solid-1/3 state -- two sublattices are fully occupied by $A$ and $B$ bosons respectively,
                           and the remaining one is empty;
\item 2Solid-2/3 state -- one sublattice is fully occupied by both $A$ and $B$ bosons,
                          and the other two are fully occupied by $A$ and $B$ bosons respectively.
\end{enumerate}
These two solid states are both of degenerate degree 6, and the internal energies are given by
\begin{eqnarray}
E_{1/3}=  -2 N \mu /3
\end{eqnarray}
and
\begin{eqnarray}
E_{2/3}= N (2 V + U/3 -4 \mu  /3),
\end{eqnarray}
where $N$ is the number of lattice sites.
The two solid states coexist at $\mu=3V+U/2$ where $E_{1/3}=E_{2/3}$.

$\texttt{Decoupled case (U=0)}$. For $U=0$, the two species are decoupled and Hamiltonian~(\ref{hamiltonian}) reduces to
\begin{eqnarray}\label{hamiltonian2}
H=\sum_{\alpha \in \{A,B \}} H_\alpha,
\end{eqnarray}
with
\begin{eqnarray}\label{hamiltonian3}
H_\alpha= -t_\alpha \! \sum_{<ij>} \alpha^+_i a_j +  V \sum_{<ij>}n^{\alpha}_i  n^{\alpha}_j
  -\mu \sum_{i}  n^{\alpha}_i.
\end{eqnarray}
Hamiltonian~(\ref{hamiltonian3}) describes a single species of hardcore bosons with nearest-neighbor repulsion $V$.
It has been extensively studied by different groups~\cite{tri2005}, and a supersolid (SS) state
was found in region near the half-integer filling ($\mu=3V$).

\section{Measured quantities}~\label{s4}

We use the worm-type quantum Monte Carlo method to simulate the Hamiltonian~(\ref{hamiltonian}).
The worm algorithm is an unbiased algorithm that works in continuous imaginary time~\cite{worm};
see Refs.~\cite{worm1} and \cite{worm2} for review. In the simulation,
the linear lattice size $L$ took several values in the range of $12 \leqslant L \leqslant 72$.
The inverse temperatures were mostly chosen as $\beta \equiv 1/T=L$,
while simulations at lower temperatures were also performed for some cases.

To explore quantum ordered phases, we measure physical quantities as:
\begin{enumerate}

\item Particle density for each species $\rho_{\alpha}=\langle  N_{\alpha}/N \rangle$,
with $N_\alpha$ the particle number of species $\alpha$ ($\alpha \in \{A,B\}$);

\item Superfluid density $\rho_{\alpha}^S=  L^{2-d} \langle W_{\alpha}^2 \rangle/\beta$,
where $W$ is the winding number~\cite{windingnumber};

\item Static structure factor
$S^{\mathbf Q}_{\alpha}=\langle \rho_{\alpha}^{\mathbf Q} \rho_{ \alpha}^{\dagger \mathbf Q} \rangle$,
where $\rho_{\alpha}^{\mathbf Q}=(1/N)\sum_i n^{\alpha}_i \exp(i {\mathbf Q} {\mathbf r_i})$ and $\mathbf Q=(4\pi/3,0)$
corresponding to $\sqrt{3}\times\sqrt{3}$ ordering;

\item CSF stiffness $\rho_{\textrm{CSF}}=L^{2-d} \langle (W_A-W_B)^2 \rangle  /\beta$,
  and PSF stiffness $\rho_{\textrm{PSF}}= L^{2-d}\langle (W_A+W_B)^2  \rangle / \beta $.

\end{enumerate}

\section{Results}~\label{s5}

\begin{figure}
\includegraphics[width=8cm,height=6.5cm]{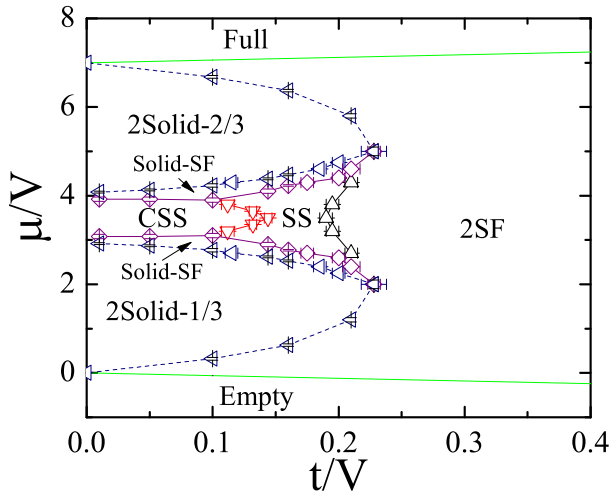}
\caption{(Color online) Phase diagram for $U/V=1$ and $t_A/V=t_B/V : \equiv t/V$.
Symbols represent critical points obtained from the simulations;
solid and dashed lines denote continuous and discontinuous phase transitions, respectively.}
\label{pd1}
\end{figure}

\begin{figure}
\centering
\includegraphics[width=8cm,height=6.5cm]{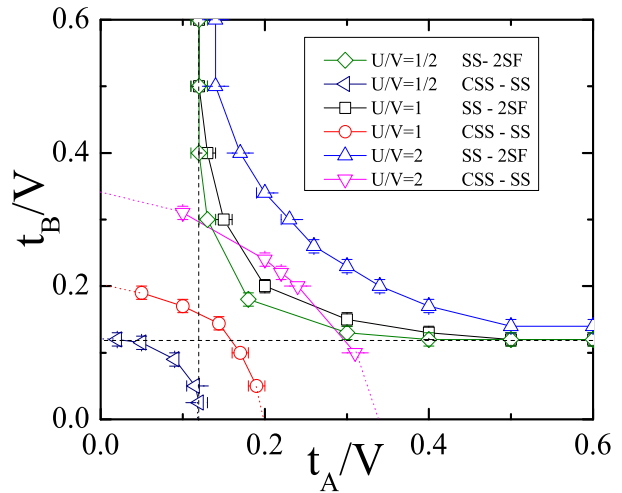}
\caption{\label{pd_all} (Color online) Phase boundaries for different $U$ values at the half-integer filling of each species.
Symbols represent critical points obtained from the simulations; solid lines denote phase boundaries. The black dashed lines
correspond to the SS-SF transitions in Model~(\ref{hamiltonian3}).}
\end{figure}

Our main findings are the phase diagrams shown in Figs.~\ref{pd1} and \ref{pd_all}.

Figure~\ref{pd1} illustrates the phase diagram for $t_A/V=t_B/V$ (say $t/V$) and $U/V=1$ on the $t/V$-$\mu/V$ plane,
which includes 2Solid-1/3, 2Solid-2/3, double SF (2SF), SS, Solid-SF, and counterflow supersolid (CSS) states.
For a given $t/V$, the lattice is empty below chemical potential $\mu=-6t$, which is the energy
to add into the lattice a pair of bosons of different species.
From the particle-hole transformation, it is known that
the phase boundaries are symmetric with respect to the line $\mu=7V/2$.

Figure~\ref{pd_all} shows the phase boundaries on the
$t_A/V$-$t_B/V$ plane with each species at the half filling, for $U/V=1/2$, $1$ and $2$,
including the CSS, SS, and 2SF states.

The quantum ordered phases can be determined by examining the robustness of measured quantities
in the limit of $L\rightarrow\infty$ and $\beta \rightarrow \infty$.
\begin{enumerate}
\item 2Solid-1/3 state --- $\forall$ $\alpha \in \{A,B\}$, $\rho_{\alpha}=1/3$, $\rho^S_{\alpha}=0$,
$S^{\mathbf Q}_{\alpha}>0$; $\rho_{\textrm{CSF}}=0$,  $\rho_{\textrm{PSF}}=0$.
\item 2Solid-2/3 state --- $\forall$ $\alpha \in \{A,B\}$, $\rho_{\alpha}=2/3$, $\rho^S_{\alpha}=0$,
$S^{\mathbf Q}_{\alpha}>0$; $\rho_{\textrm{CSF}}=0$,  $\rho_{\textrm{PSF}}=0$.
\item 2SF state ---   $\forall$ $\alpha \in \{A,B\}$, $\rho^S_{\alpha}>0$,  $S^{\mathbf Q}_{\alpha}=0$; $\rho_{\textrm{CSF}}>0$,  $\rho_{\textrm{PSF}}>0$.
\item CSS state ---   $\forall$ $\alpha \in \{A,B\}$, $\rho^S_{\alpha}>0$,  $S^{\mathbf Q}_{\alpha}>0$; $\rho_{\textrm{CSF}}>0$,  $\rho_{\textrm{PSF}}=0$.
\item SS state ---  $\forall$ $\alpha \in \{A,B\}$, $\rho^S_{\alpha}>0$,  $S^{\mathbf Q}_{\alpha}>0$; $\rho_{\textrm{CSF}}>0$,  $\rho_{\textrm{PSF}}>0$.
\item Solid-SF state ---  $\rho^S_{A}=0$, $\rho^S_{B}>0$ or $\rho^S_{A}>0$, $\rho^S_{B}=0$; $\forall$ $\alpha \in \{A,B\}$, $S^{\mathbf Q}_{\alpha}>0$; $\rho_{\textrm{CSF}}>0$, $\rho_{\textrm{PSF}}>0$.
\end{enumerate}
To further explore these quantum phases, we also took snapshots of world-line configurations
and performed histogram analyses.

In comparison with the classical limit which only exhibits the 2Solid-1/3 and 2Solid-2/3 states, quantum fluctuations
lead to many unusual states that break a variety of symmetries.
In the 2SF state, off-diagonal superfluid order develops for each species, and
an $U(1) \times U(1)$ symmetry is broken. The Solid-SF, SS and CSS states exhibit
simultaneously broken translational and $U(1)$ symmetries, as well as broken symmetry between the two species.

The following subsections present numerical evidences for the aforementioned findings.
Subsection $\mathbf{A}$ constructs the phase diagram in Fig.~\ref{pd1} for $t_A/V=t_B/V$ and $U/V=1$.
Subsection $\mathbf{B}$ determines the phase boundaries in Fig.~\ref{pd_all} at half-integer fillings
and reveals the effects induced by the variation of $U/V$.

\subsection{$t_A/V=t_B/V$ and $U/V=1$}~\label{ss1}

\begin{figure}
\includegraphics[width=8cm,height=9cm]{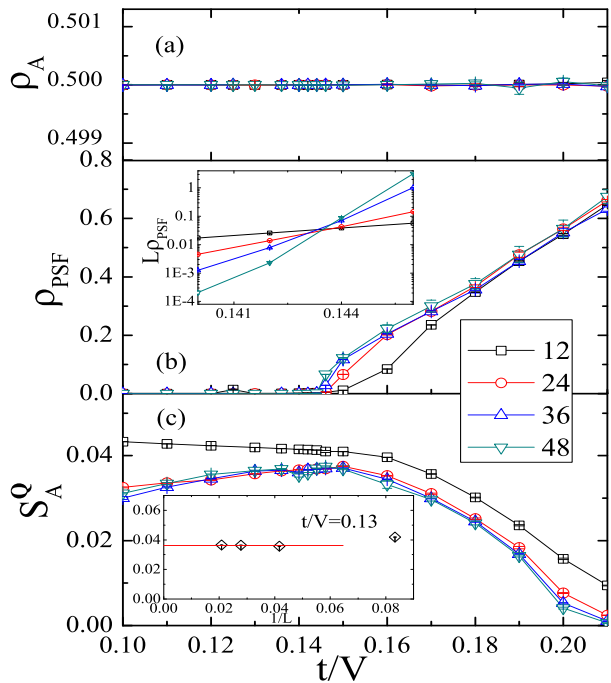}
\caption{(Color online) Quantities $\rho_A$ (a), $\rho_{\textrm{PSF}}$ (b) and $S^{\mathbf Q}_A$ (c) versus $t/V$ at $\mu/V=3.5$ and $U/V=1$.}
\label{mu35}
\end{figure}

$\texttt{Half-integer fillings}$. We perform grand-canonical simulations with $\mu/V = 7/2$, which yield particle density
$\rho_A=\rho_B=0.5000(1)$ [see Fig.~\ref{mu35}(a)].

As hopping amplitude $t$ increases, the PSF stiffness $\rho_{\textrm {PSF}}$ starts
to become nonzero near $t/V \sim 0.15$, as shown in Fig.~\ref{mu35}(b).
To more precisely locate the phase transition point, we perform a finite-size scaling analysis of the $\rho_{\textrm {PSF}}$ data.
At the transition point, it is expected that the PSF stiffness scales as $\rho_{\textrm {PSF}}=L^{2-d-z}f(\beta/L^{z})$,
where dynamical critical exponent $z$ equals to $1$ if the system has the particle-hole symmetry.
The inset of Fig.~\ref{mu35}(b) plots the scaled PSF stiffness $L\rho_{\textrm {PSF}}$ versus $t/V$,
which yields the transition point as $t/V=0.144(2)$ from the approximate common intersection for different $L$.
The $S_A^{\mathbf Q}$ data in Fig.~\ref{mu35}(c) suggest another transition point at $t/V=0.20(1)$ beyond which
the crystalline order vanishes.
These two transition points should separate three phases, which shall be identified below.

\begin{figure}
\begin{center}
\includegraphics[width=8cm,height=5cm]{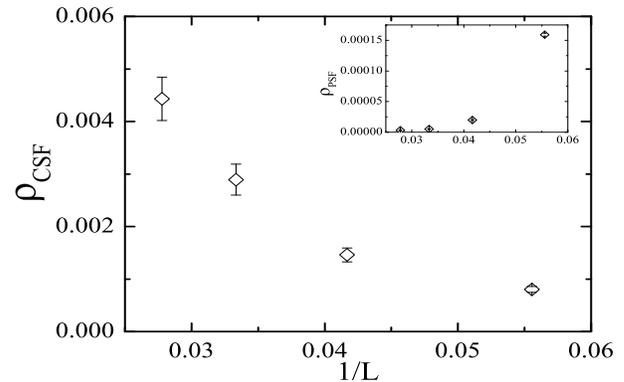}
\caption{Quantity $\rho_{\textrm {CSF}}$  versus $1/L$ at $\mu/V=3.5$, $t/V=0.13$ and $U/V=1$.
Inset: $\rho_{\textrm PSF}$ versus $1/L$.
For these two figures, simulations are performed with $\beta=5L$ (see text).}
\label{ps_scf}
\end{center}
\end{figure}

We first present numerical evidence for the CSS state in the region $t/V < 0.144(2)$.
The nonzero crystalline order for each species is demonstrated by Fig.~\ref{mu35}(c)
and by the inset which plots $S_A^{\mathbf Q}$ versus $1/L$ for $t/V=0.13$.
It is known~\cite{BCS} that the CSF density $\rho_{\textrm {CSF}}$ is fragile against thermal fluctuations.
Thus, to detect the CSS order, we simulate at temperature $\beta=5L$ for $t/V=0.13$.
As shown in Fig.~\ref{ps_scf} and the inset, the CSF density $\rho_{\textrm {CSF}}$
clearly approaches a nonzero value as $L$ increases;
in contrast, the PSF density $\rho_{\textrm {PSF}}$ drops drastically to zero.
These evidences establish the CSS state for $t/V < 0.144(2)$.
A histogram analysis is performed to further explore the particle-number distribution.
In addition to the broken $U(1)$ symmetry due to the long-range off-diagonal order $\rho_{\textrm {CSF}}$,
it is observed that both the symmetry among the three sublattices
and the symmetry between the two species are spontaneously broken.
Namely, the particle density $\rho_{\alpha,s}$ on sublattice $s$ for species $\alpha$
can vary for different species and for different sublattices--(say $S_1$, $S_2$, $S_3$).
There are 6-fold ground states, in one of which the filling factor is arranged as ($1$,$\frac{1}{4}$,$\frac{1}{4}$) for $A$ bosons
and ($0$,$\frac{3}{4}$,$\frac{3}{4}$) for $B$ bosons.
The bosons on sublattice $S_1$ are pinned, and the counter-flow superfluidity arises
from $S_2$ and $S_3$, which form a honeycomb lattice.
Finally, we note that despite of symmetry breaking between $A$ and $B$ bosons,
the total particle numbers are identical for the $A$ and $B$ bosons--i.e., $\sum_s \rho_{\alpha,s}=3/2$,
and the summed filling factor of the two species is unity for each sublattice--i.e., $\sum_\alpha \rho_{\alpha,s}=1$.

\begin{figure}
\begin{center}
\includegraphics[height=3.0cm,width=0.55\columnwidth]{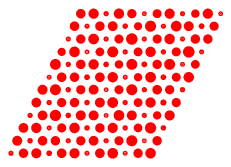}
\hspace*{-12mm}
\includegraphics[height=3.0cm,width=0.55\columnwidth]{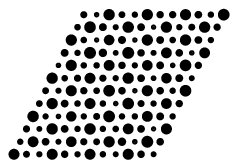}
\end{center}
 \caption{\label{175} (Color online) Typical particle distributions on a world-line configuration in the SS state ($t/V=0.175$, $\mu/V=3.5$, and $U/V=1$) for $A$ (left) and $B$ (right) species respectively. This world-line configuration is obtained after sufficient Monte Carlo steps to achieve equilibrium. The simulation is on a $36 \times 36$ lattice, but for illustrating purpose, we show a block of $12 \times 12$ sites.}
\end{figure}

In the region $0.144(2) < t/V < 0.20(1)$, the system is in the SS state and
characterized by nonzero crystalline order and nonzero superfluid density for both species,
as demonstrated in Figs.~\ref{mu35}(b) and (c).
In comparison with the CSS state, the degenerate degree of the ground state is also $6$
but the particle distribution is distinct, arranged as ($\frac{1}{6}$,$\frac{2}{3}$,$\frac{2}{3}$) for
one species and ($\frac{5}{6}$,$\frac{1}{3}$,$\frac{1}{3}$) for the other.
Also note that the two species have equal total numbers of bosons and
the summed filling factor on each sublattice is unity.
The bosons on the sublattice of filling factor $\left( \frac{1}{6}+\frac{5}{6} \right)$
are pinned while those on the remaining two sublattices account for the superfluidity.
Figure~\ref{175} illustrates a snapshot of particle number distribution for $t/V=0.175$ and $L=36$,
which is averaged over the imaginary-time axis of a world-line configuration.

For $t/V > 0.20(1)$, the system is in a 2SF state, featured by nonzero superfluid density
and zero crystalline order for both species.

\begin{figure}
\includegraphics[width=8cm,height=9cm]{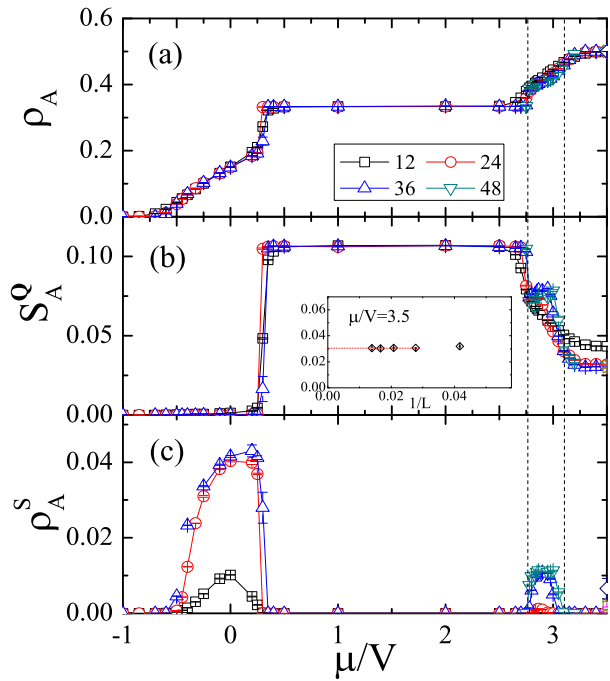}
\caption{(Color online) Quantities $\rho_A$ (a), $S^{\mathbf Q}_A$ (b) and $\rho^S_A$ (c) versus $\mu/V$ for $t/V=0.1$ and $U/V=1$. The region between
  the two dashed lines corresponds to Solid-SF state of which the data are average results of Monte Carlo simulations with different initial conditions.}
\label{t01}
\end{figure}

\begin{figure}
\includegraphics[width=8cm,height=4cm]{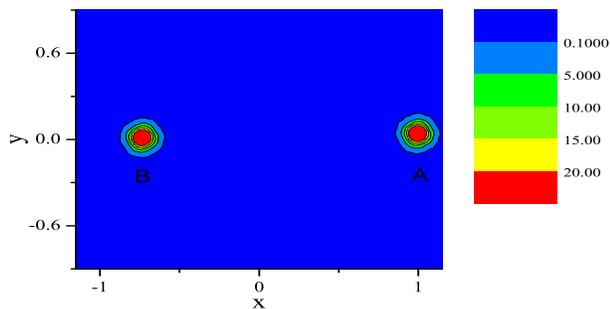}
\caption{(Color online) Histogram of quantities $\vec{n}_{A}$ and $\vec{n}_{B}$ obtained from a single simulation in the Solid-SF state.  The simulation is for the parameter set $t/V=0.1$, $\mu/V=2.9$, $U/V=1$, $L=36$ and $\beta=720$, and determines $\rho_A=0.333(1)$ and $\rho_B=0.480(1)$.
The determined particle density of $A$ species consists with the expected value $1/3$ (with a tiny relative error),
providing another evidence supporting that $A$ species forms a commensurate solid.}
\label{SSSolid}
\end{figure}

\begin{figure}
%\vspace*{0cm} \hspace*{-0cm}
\begin{center}
%%\quad \vspace{4cm}  %% TEMPORARY UNTIL FILE IS THERE
\includegraphics[height=3.0cm,width=0.55\columnwidth]{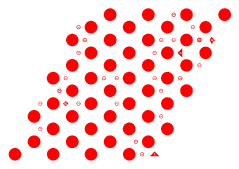}
\hspace*{-12mm}
\includegraphics[height=3.0cm,width=0.55\columnwidth]{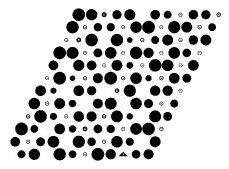}
\end{center}
 \caption{\label{01conf}(Color online)  Typical particle distribution on a world-line configuration in the Solid-SF state ($t/V=0.1$, $\mu/V=2.9$, and $U/V=1$) for $A$ (left) and $B$ (right) species respectively. This world-line configuration is obtained after sufficient Monte Carlo steps to achieve equilibrium. The simulation is on a $36 \times 36$ lattice, but for illustrating purpose, we show a block of $12 \times 12$ sites.}
\end{figure}

$\texttt{t/V=0.1}$.
We describe the simulations for $U/V=1$ and away from the symmetric line $\mu/V=7/2$
in the example of $t/V=0.1$ with varying $\mu/V$.
As the chemical potential $\mu$ increases, the system in the 2SF phase is driven into the 2Solid-1/3 state
by a first-order phase transition at $\mu/V=0.32(3)$,
reflected by the discontinuities of $\rho_A$, $\rho^S_A$ and $S_A^{\mathbf Q}$ in Fig.~\ref{t01}.

In the region $2.77(2)< \mu/V< 3.10(5)$, the system is detected to be in a novel quantum state
which exhibits nonzero crystalline order for both species but nonzero superfluid density for one species only.
The ground state is still 6-fold, and the particle distribution on sublattice ($S_1$,$S_2$,$S_3$) is arranged as
(1,0,0) for one species--say $A$ bosons--and (0,$\frac{3\rho_B}{2}$,$\frac{3\rho_B}{2}$) for $B$ bosons,
where $\rho_B$ is a function of $\mu/V$.
The $A$ bosons are pinned on sublattice $S_1$, and do not exhibit visible superfluid density.
The $B$ bosons are only present on sublattices $S_2$ and $S_3$, and contribute to a nonzero superfluid density.
Unlike in the CSS and SS states, the total particle numbers of the $A$ and $B$ bosons
are no longer identical; one has $\rho_A = (1/3)\sum_s \rho_{s,A} = 1/3$ while $\rho_B = (1/3)\sum_s \rho_{s,B}$
varies with $\mu/V$.
We provide further analyses by simulating at a rather low temperature $\beta = 720$ for $(\mu/V=2.9, L=36)$.
We define a vector order parameter for each species
$\vec{n}_{\alpha}=\rho_{1,\alpha}+\rho_{2,\alpha} e^{i 2\pi/3}+\rho_{3,\alpha} e^{i4\pi/3}$ ($\alpha=A$ and $B$).
The histogram is shown in Fig.~\ref{SSSolid}, where the probability distribution of $\vec{n}_A$
is around point (1,0) and that of $\vec{n}_B$ is around point (-$\frac{3\rho_B}{2}$,0) with $\rho_B=0.480(1)$.
The imbalance of the total particle numbers of the two species is clearly seen.
A snapshot of the particle distribution is shown in Fig.~\ref{01conf}.
Nonzero superfluid density $\rho_B^S$ is observed,
while $\rho_A^S$ is $0.0000(1)$, consistent with zero. Note that the density wave of $B$ species is induced by $A$ species which forms an insulating solid. Therefore, $B$ species should be known as SF rather than SS, and the mixture can be called Solid-SF.
This treatment was also discussed in Ref.~\cite{TC1}.
We conclude this paragraph by mentioning the following.
In Ref.~\cite{Trousselet2013}, a similar phase was observed away from the half-filling case
in parameter region including ($U/V =2, t/V=0.15$),
which is characterized by a 3-fold order for both species and a nonzero superfluid density
for one species only. This phase was referred to the supersolid/3-fold-order (SS/3FO) phase.
However, according to Ref.~\cite{Trousselet2013}, the SS/3FO phase has rather distinct particle distribution:
two sublattices are respectively filled by the two species while on the remaining, a species of bosons
accounts for the superfluidity.
This is inconsistent with our observation for the Solid+SF phase, in which the superfluidity arises
from one species of bosons on two sublattices that form a honeycomb lattice.

For the region $3.10(5) < \mu/V \leq 3.5$, the system is in the CSS state, supported by
a robust crystalline order for each species together with
vanishing pair superfluidity and nonzero counterflow superfluidity.

The whole phase diagram in Fig.~\ref{pd1} is constructed by simulations with a variety of ($t/V$, $\mu/V$) values.

\subsection{\rm $t_A/V \neq t_B/V$ and/or $U/V \neq 1$}~\label{ss2}

In this subsection, we shall study the effects induced by the asymmetry between the $A$ and $B$ bosons
due to imbalanced hopping amplitudes $t_A/V \neq t_B/V$.
The effects by tuning interaction $U/V$ are also considered.

\texttt{$U/V = 1$ and $\mu/V=7/2$}.  We first study the $t_A/V \neq t_B/V$ effect for ($U/V=1$, $\mu/V=7/2$),
with each species at the half-integer filling.

\begin{figure}
\includegraphics[width=8cm,height=9cm]{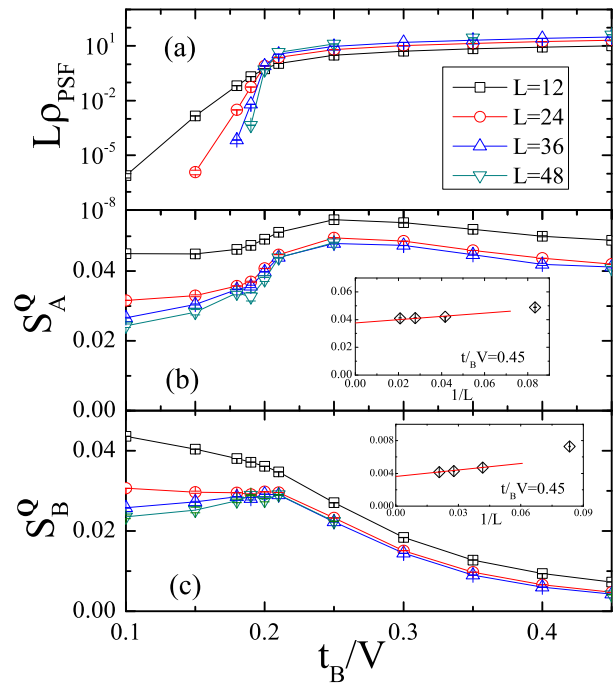}
\caption{(Color online) Quantities $L\rho_{\textrm {PSF}}$ (a), $S^{\mathbf Q}_A$ (b) and $S^{\mathbf Q}_B$ (c) versus  $t_B/V$ at $t_A/V=0.05$ and $U/V=1$.}
\label{ani_t005}
\end{figure}

We simulate at $t_A/V=0.05$ with varying $t_B/V$,
and Fig.~\ref{ani_t005}(a) shows $L\rho_{\textrm {PSF}}$ for different $L$,
indicating a continuous phase transition at $t_B/V=0.19(1)$.
Near this point, a kink is observed in $S_A^{\mathbf Q}$ [Fig.~\ref{ani_t005}(b)].
Nevertheless, no sharp decrease of $S_A^{\mathbf Q}$ exists in either side of the transition;
actually, $S_A^{\mathbf Q}$ in the whole $t_B/V$ range converges to nonzero values as $L \rightarrow \infty$.
A finite-size analysis is shown in the inset of Fig.~\ref{ani_t005}(b) for $t/V=0.45$.
Similar feature is found for $S_B^{\mathbf Q}$ [Fig.~\ref{ani_t005}(c)].
This means that both species exhibit a crystalline order in the whole $t_B/V$ range.
Together with the behavior of $\rho_{\textrm {CSF}}$ (not shown), it can be established that
the system is in the CSS state for $t_B/V<0.19(1)$ and the SS state for $t_B/V > 0.19(1)$.

It is interesting to note that as $t_B/V$ increases, the crystalline order of $B$ bosons is not destroyed.
The underlying reason is that such an order is induced by the translational-symmetry breaking
due to $A$ bosons which are in the SS state.

Simulations have been carried out for a variety of $t_A/V$ values for $(U/V=1, \mu/V=7/2)$.
The phase diagram in the $(t_A/V, t_B/V)$ plane (Fig.~\ref{pd_all}) contains the CSS, SS, and 2SF states.

$\texttt{U/V varies}$. To study the effects induced by the variation of interaction ratio $U/V$,
we simulate for $U/V=1/2$ and $2$ at the half-integer filling ($\mu=3V+U/2$).
The phase diagrams (Fig.~\ref{pd_all}) for different $U/V$ are of similar topology,
and contain two phase boundaries separating the CSS, SS and 2SF states.
As $U/V$ increases, the CSS region gets broader, while
the SS phase near the symmetric line $t_A/V=t_B/V$ drastically shrinks.
As $U/V$ decreases, the SS-2SF phase boundary gets closer and closer to $t_A/V\approx 0.12$ or $t_B/V\approx 0.12$
which are the SS-SF transition points of Model~(\ref{hamiltonian3})
(denoted as black dashed lines in Fig.~\ref{pd_all})~\cite{tri2005}.
The SS state is persistent up to asymmetric hopping limits
($t_A/V \rightarrow\infty $, $t_B/V \lessapprox 0.12$) and ($t_A/V \lessapprox 0.12$, $t_B/V \rightarrow\infty$).

\section{Discussions}~\label{s6}
We have explored the quantum ordered phases in bosonic mixtures on the triangular lattice,
which are constituted by two species of hardcore bosons, by using extensive Monte Carlo simulations.
These quantum ordered phases are determined by complementary approaches: examining the
robustness of measured quantities, analyzing world-line configurations, and performing histogram analyses.
For $t_A/V=t_B/V$ (say $t/V$) with $U/V=1$, we constructed a complete ground-state phase diagram (Fig.~\ref{pd1}) on
the $t/V$-$\mu/V$ plane, which includes 2Solid-1/3, 2Solid-2/3, 2SF, SS, Solid-SF and CSS states.
We then considered the cases with $t_A/V \neq t_B/V$ and found that the SS and CSS states are present
in a broad parameter range (Fig.~\ref{pd_all}).
Further, the SS state even survives up to asymmetric limits of hopping amplitudes.
We also explored the effects induced by the variation of the interaction ratio $U/V$ which is experimentally tunable.

In the CSS, SS and Solid-SF states, both the symmetry among the three sublattices and the symmetry between the two species
are spontaneously broken, and the ground state is 6-fold.
To further check the robustness of these states and the 6-fold degeneracy,
we slightly break the balance of the chemical potentials of the two species
such that $\mu_A=\mu+0^+$ and $\mu_B=\mu+0^-$.
No qualitative change is observed in the CSS state, but the SS and Solid-SF states both become 3-fold.
In the SS state, $A$ bosons prefer to filling arrangement $(\frac{1+0^+}{6}, \frac{2+0^+}{3}, \frac{2+0^+}{3})$
while $B$ species is of the $(\frac{5+0^-}{6}, \frac{1+0^-}{3}, \frac{1+0^-}{3})$ structure.
In the Solid-SF phase with $\mu<3V+U/2$, the $B$ species is in the solid state with filling factor $(1,0,0)$,
while the $A$ bosons are in the superfluid phase.
This demonstrates that the whole region of the SS and Solid-SF states is a surface of
first-order phase transitions.

When finishing most of the Monte Carlo simulations for this manuscript,
we became aware of the recent work by Trousselet $et$ $al.$~\cite{Trousselet2013}
who studied the same system with a mean-field approach and exact diagonalizations.
While most of the quantum phases in our work have already been predicted in Ref.~\cite{Trousselet2013},
quantitative difference does exist in the location of phase boundaries.
Furthermore, via snapshots of world-line configuration and histogram analyses,
this work provides strong and direct evidence for the CSS, SS, and Solid-SF phases,
which are respectively termed as ``color supersolid", ``bosonic pinball",
and ``SS/3FO" phases in Ref.~\cite{Trousselet2013}.
Finally, the filling factors in the Solid-SF phase are found to be qualitatively different.

Experimental studies of bosonic mixtures on the triangular optical lattices are called
for to test the novel quantum phases in this work and in Ref.~\cite{Trousselet2013}.
Since the present model has many tunable parameters,
other non-trivial quantum ordered states which are located at certain parameter ranges
may be still uncovered, which request further theoretical work.
For example, pair-supersolid state may emerge if the onsite interspecies
interaction becomes attractive ($U<0$)~\cite{Trefzger2009}. The present algorithm is potentially
applicable in some spin-boson models such as those in Refs.~\cite{Winter} and ~\cite{Zhang}.

\begin{acknowledgments}
The manuscript received great help from Nikolay Prokof'ev and Boris Svistunov,
who shared code with us, made critical reading of the manuscript, and provided insightful suggestions.
This work was supported by the National Basic
Research Program of China (Grant Nos. 2011CBA00103 and 2011CB921304),
National Natural Science Foundation of China (Grant Nos. 11147013 and 11275185),
and the Fundamental Research Funds for the Central Universities (Grant Nos. 2012QNA43 and 2011RC25).
\end{acknowledgments}

\end{document}